\documentclass[amsmath,amssymb,prd, notitlepage,nofootinbib,twocolumn, superscriptaddress]{revtex4-2}

\usepackage{graphicx}
\usepackage{dcolumn}
\usepackage{bm}
\usepackage{float}
\usepackage{xcolor}
\usepackage{comment}
\usepackage[caption=false]{subfig}
\usepackage{fontawesome5}
\usepackage{hyperref}
\usepackage[normalem]{ulem}
\newcommand{\mnras}{Monthly Notices of the Royal Astronomical Society}
\newcommand{\jcap}{Journal of Cosmology and Astroparticle Physics}

\newcommand{\github}[1]{%
   \href{#1}{\faGithubSquare}%
}

\begin{document}

\preprint{APS/123-QED}

\title{Constraints on Dark Matter-Dark Energy Scattering from ACT DR6 CMB Lensing}
\author{Alex Lagu\"e}%
\email{alague@sas.upenn.edu}
\affiliation{Department of Physics and Astronomy, University of Pennsylvania, 209 South 33rd Street, Philadelphia, PA, USA 19104}%

\author{Fiona McCarthy}
\affiliation{DAMTP, Centre for Mathematical Sciences, Wilberforce Road, Cambridge CB3 0WA, UK}
\affiliation{Kavli Institute for Cosmology Cambridge, Madingley Road, Cambridge, CB3 0HA, UK}
\affiliation{Center for Computational Astrophysics, Flatiron Institute, 162 5th Avenue, New York, NY 10010 USA}


\author{Mathew Madhavacheril}
\affiliation{Department of Physics and Astronomy, University of Pennsylvania, 209 South 33rd Street, Philadelphia, PA, USA 19104}%
\author{J.~Colin Hill}
\affiliation{Department of Physics, Columbia University, New York, NY, USA 10027}%
\author{Frank J. Qu}
\affiliation{DAMTP, Centre for Mathematical Sciences, Wilberforce Road, Cambridge CB3 0WA, UK}
\affiliation{Kavli Institute for Cosmology Cambridge, Madingley Road, Cambridge, CB3 0HA, UK}

\date{\today}

\begin{abstract}
The predicted present-day amplitude of matter fluctuations based on cosmic microwave background (CMB) anisotropy data has sometimes been found  discrepant with more direct measurements of late-time structure. This has motivated many extensions to the standard cosmological model, including kinetic interactions between dark matter and dark energy that introduce a drag force slowing the growth
of structure at late times. Exploring this scenario, we develop a model for quasi-linear scales in the matter power spectrum by calculating the critical overdensity in the presence of this interaction and a varying dark energy equation of state. We explicitly avoid modeling or interpretation of data on non-linear scales in this model (such as use of $\Lambda$CDM-calibrated priors), which would require numerical simulations. 
We find that the presence of the drag force hinders halo formation, thus increasing the deviation from $\Lambda$CDM in the mildly non-linear regime. We use CMB lensing observations from the sixth data release of the Atacama Cosmology Telescope up to $L=1250$ (in combination with \textit{Planck}, Sloan Digital Sky Survey, and  6dFGS data) to derive the strongest constraints to date on the amplitude of the drag term, finding the dimensionless interaction strength $\Gamma_\mathrm{DMDE}/(H_0\rho_\mathrm{c})<0.831\; (2.81)$ at the 68\% (95\%) confidence level. The inclusion of non-linear corrections improves our constraints by about 35\% compared to linear theory. Our results do not exclude the best-fit values of $\Gamma_\mathrm{DMDE}$ found in previous studies using information from galaxy weak lensing, though we find no statistical preference for the dark matter-dark energy kinetic interactions over $\Lambda$CDM. We implement our model in a publicly available fork of the Boltzmann code \texttt{CLASS} at \github{https://github.com/fmccarthy/Class_DMDE}.

\end{abstract}

\maketitle



\section{\label{sec:intro} Introduction}
Some recent measurements of the amount of clustering and matter content in the Universe obtained using the cosmic microwave background (CMB) and weak gravitational lensing from nearby galaxies have yielded conflicting results. To quantify this discrepancy, we use the variance of the linear-theory density field smoothed on a scale of 8 Mpc/$h$ ($\sigma_8$) rescaled by the total matter density at the present day: $S_8\equiv \sigma_8 (\Omega_\mathrm{m}/0.3)^{0.5}$. The variance of the matter density can be calculated as an integral over the linear power spectrum
\begin{align}
    \sigma^2_8(z) &\equiv \sigma^2(R=8.0\;h/\mathrm{Mpc};z) \\&= \int \frac{dk}{k} \frac{k^3 P(k, z)}{2\pi^2} W^2(kR),
\end{align}
where $W$ is the Fourier transform of a spherical top-hat filter. The present-day value of $S_8$ can be indirectly inferred from an extrapolation of observations of primary CMB anisotropies that capture the early Universe around recombination. Various CMB experiments broadly agree on this measurement: the \textit{Planck} satellite 2018 release finds $S_8 = 0.830 \pm 0.013$~\cite{Planck2020Parameters}, while the combination of {\it WMAP} and the Atacama Cosmology Telescope (ACT) DR4 finds $S_8= 0.840 \pm 0.030$~\cite{Aiola2020TheAtacama}.  Meanwhile, weak lensing measurements from galaxy surveys that measure the clustering of matter directly at later times  give somewhat lower values of $S_8 =0.763^{+0.040}_{-0.036}$ for the Hyper Suprime-Cam Y3~\cite{Miyatake2023HSC-YR3}, $S_8 =0.776\pm {0.017}$ for the Dark Energy Survey Year 3~\cite{DES2022YR3Params}, and $S_8 =0.773^{+0.028}_{-0.030}$ for the Kilo-Degree Survey~\cite{Dvornik2023KiDS-1000}, among others (although a joint analysis of the Kilo-Degree Survey and the Dark Energy Survey found a value closer to the primary CMB results with $S_8=0.790^{+0.018}_{-0.014}$~\cite{DESKiDS2023ConsistentCosmology}). Intriguingly, gravitational lensing of the CMB (probing larger scales and intermediate cosmic times) yields $S_8 = 0.840 \pm 0.028$ for the Atacama Cosmology Telescope (ACT DR6)~\cite{Qu2023ACTLensing,Madhavacheril2023ACTLensing}, $S_8 = 0.831 \pm 0.029$ for \textit{Planck} PR4 lensing (NPIPE)~\cite{Carron2022CMBLensing}, while SPT-3G lensing finds $S_8 = 0.836 \pm 0.039$~\cite{Pan2023MeasurementOf}, in very good agreement with the primary CMB extrapolations. 

These values are obtained assuming a flat $\Lambda$CDM model of the Universe. Many modifications to this model have been proposed to alleviate discrepancies between $S_8$ measurements. Most of them fall into three broad categories: (a) models that change the shape of the matter power spectrum on small scales, (b) models that modify the growth of structure over time, and (c) models that do a combination of both. In the present work, we constrain a model in the third category that predicts a slower growth of structure on (predominantly) small scales at late times.

The model we consider in this study allows for a transfer of momentum between the dark matter and dark energy components parameterized by a scattering rate: $\Gamma_{\rm DMDE}$~\cite{Simpson2010ScatteringOf, Poulin2023Sigma-8}. It is a two-parameter extension of $\Lambda$CDM where the dark energy equation of state can vary from the cosmological constant value of $w=-1$. When the scattering rate is non-zero, there is an additional friction or `drag' term in the equation of motion of the matter density perturbations, which leads to a highly redshift-dependent suppression of the growth of structure. This model has been constrained previously using data from the primary CMB, baryon acoustic oscillations (BAO), and galaxy weak lensing~\cite{Kumar2017ObservationalConstraints,Asghari2019OnStructure,Poulin2023Sigma-8}, though these works used exclusively linear theory. The accuracy of this  approximation degrades on small scales and could bias the value of the interaction rate inferred from the data as well as affect the preference for this model over $\Lambda$CDM. In this article, we address these concerns by (a) developing a halo model to compute the matter power spectrum on weakly non-linear scales and (b) using lensing of the CMB, which is sensitive to larger scales and higher redshifts than galaxy weak lensing~\cite{Qu2023ACTLensing}.

\begin{figure}[h]
    \centering
    \includegraphics[width=0.95\linewidth]{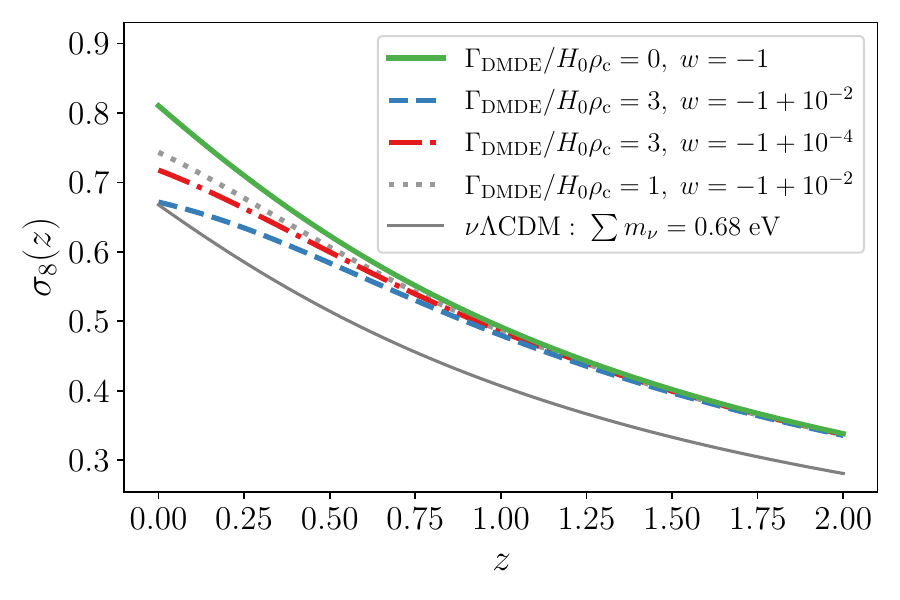}
    \caption{Impact of DM-DE scattering on $\sigma_8$ as a function of redshift and DE equation of state. The thin grey line has the same cosmology as the green curve, but with a sum of neutrino masses set to match the $\sigma_8(z=0)$ of the dashed blue curve. This demonstrates the difference in redshift evolution between $\nu\Lambda$CDM and $w\Gamma$CDM.}
    \label{fig:sig8}
\end{figure}

This paper is organized as follows. First, in Sec.~\ref{sec:theory}, we present the theoretical background for the model and its numerical implementation. In Sec.~\ref{sec:data} we list the dataset combinations we use to constrain our model parameters, and in Sec.~\ref{sec:results} we present the results of our statistical analysis. Finally, in Sec.~\ref{sec:discussion}, we discuss the implications of our findings and compare our results with those of previous analyses.


\section{ Theory\label{sec:theory}} 

\subsection{Linear Perturbations}

We consider a kinetic interaction between dark matter (DM) and dark energy (DE), which couples their velocity divergences. This DM-DE scattering model, which we refer to as $w\Gamma$CDM, was originally constructed in Ref.~\cite{Simpson2010ScatteringOf}. It was investigated more recently in Ref.~\cite{Poulin2023Sigma-8}. We assume that the DE component has an equation of state parameter $w> -1$ and a sound speed $c_s^2=1$ (we use units with the speed of light $c=1$ throughout this paper). 
In this case, the perturbation equations for the DM and DE components in the Newtonian gauge read~\cite{Ma1995CosmologicalPerturbation, Simpson2010ScatteringOf} 
\begin{align}
    \delta_{\mathrm{DM}}^{\prime} &= -\theta_{\mathrm{DM}} + 3\phi^{\prime},\label{eq:DM-dens-deriv}\\
    \delta_{\rm DE}^{\prime} & =-\left[(1+w)+9 \frac{\mathcal{H}^2}{k^2}\left(1-w^2\right)\right] \theta_{\rm DE}\nonumber \\ &\;\;\;\;+3(1+w) \phi^{\prime}-3 \mathcal{H}(1-w) \delta_{\rm DE}, \label{eq:DE-dens-deriv}\\
    \theta_{\mathrm{DM}}^{\prime} &= -\mathcal{H} \theta_{\mathrm{DM}}+k^2 \phi+\frac{a \Gamma_{\rm DMDE}}{\bar{\rho}_{\rm DM}}\Delta \theta, \label{eq:DM-vel-deriv}\\
    \theta_{\mathrm{DE}}^{\prime}= & 2 \mathcal{H} \theta_{\mathrm{DE}}+\frac{k^2}{\left(1+w\right)} \delta_{\mathrm{DE}} +k^2 \phi-\frac{a\Gamma_{\mathrm{DMDE}}}{(1+w)\bar{\rho}_{\rm DE}} \Delta \theta \label{eq:DE-vel-deriv},
\end{align}
where $\delta\equiv\rho/\bar{\rho}-1$ is the energy density perturbation, $\theta\equiv \nabla\cdot \mathbf{v}$ is the velocity divergence, $\Delta \theta \equiv \theta_{\rm DE}-\theta_{\rm DM}$,  $\mathcal H$ is the conformal Hubble factor, $\phi$ is the Newtonian gravitational potential, and $\Gamma_{\rm DMDE}$ is a constant parameter proportional to the DM-DE cross-section which quantifies the strength of this interaction rate~\cite{Simpson2010ScatteringOf}. The bars denote spatial averages and the primes denote differentiation with respect to conformal time. The system of equations (\ref{eq:DM-dens-deriv})-(\ref{eq:DE-vel-deriv}) describes the evolution of the DM and DE perturbations. In the $\Lambda$CDM model, there are no DE perturbations, and in models where $\delta_{\rm DE}$ is allowed to grow, the coupling to DM happens through the gravitational potential $\phi$. In the case of the DM-DE drag force, the coupling between the DE and DM equations happens through the difference in their respective velocities. Multiple models explore dynamical couplings between DM and DE using field theory approaches~\cite{Pourtsidou2016ReconcilingCMB} or numerical simulations of microphysical models~\cite{Baldi2015SimulatingMomentum,Baldi2017StructureFormation}. We modify the Boltzmann code \texttt{CLASS}~\cite{Blas2011CLASS} to incorporate the $\Gamma_{\rm DMDE}$ parameter (the \texttt{CLASS} version developed in this work is publicly available~\footnote{\url{https://github.com/fmccarthy/Class_DMDE}}).

In the sub-horizon limit, the dark energy perturbations do not grow given their sound speed. In this case, the DM perturbations obey
\begin{align}
    \ddot{\delta}_{\rm DM} + 2H\left(1+\frac{\Gamma_{\rm DMDE}}{2H\bar{\rho}_{\rm DM}}\right)\dot{\delta}_{\rm DM} - 4\pi G \bar{\rho}_{\rm DM} \delta_{\rm DM} = 0,\label{eq:main-linear-delta}
\end{align}
where the overdot denotes differentiation with respect to time. Eq.~(\ref{eq:main-linear-delta}) is the non-relativistic (i.e. sub-horizon) version of the system of equations (\ref{eq:DM-dens-deriv})-(\ref{eq:DE-vel-deriv}) and describes the evolution of DM perturbations in the Newtonian limit. In $\Lambda$CDM, the perturbations behave as a damped harmonic oscillator where the friction force is given by the rate of expansion of the Universe. We note that the role of $\Gamma_{\rm DMDE}$ is to increase the amplitude of the friction term in the evolution of DM perturbations, thus slowing their growth. We recover the differential equation for the usual cold DM perturbations when $\Gamma_{\rm DMDE}\to 0$. Since $\bar{\rho}_{\rm DM} \propto a^{-3}$, the impact of the DM-DE scattering is highly redshift-dependent. At late times, we have $a\sim 1$ and $\Gamma_{\rm DMDE}/(2H\bar{\rho}_{\rm DM}) \sim \Gamma_{\rm DMDE}/(2\Omega_{\rm m} H_0\rho_{c}) \sim \Gamma_{\rm DMDE}/(H_0\rho_{c})$ so we expect deviations from $\Lambda$CDM to be significant for $\Gamma_{\rm DMDE}/(H_0\rho_{c})$ of order one\footnote{These order-of-magnitude approximations are valid given $\rho_{\rm DM} \approx \rho_{\rm m}=\Omega_{\rm m}\rho_c$ and from $2\Omega_{\rm m}\sim \mathcal{O}(1)$.}. Importantly, as argued in Ref.~\cite{Asghari2019OnStructure}, the suppression happens on sub-horizon scales and does not affect the background dynamics (indeed, the DM-DE coupling is a pure momentum exchange, and thus only appears at first order in perturbations). Thus any change in the expansion history compared to $\Lambda$CDM (which affects the redshift where DE starts to dominate) is due to varying $w$ away from $-1$. This allows the DM-DE model to suppress the growth of structure without lowering $\Omega_{\rm m}$. Furthermore, while the coupling introduces further acoustic oscillations, these are more present in the DE and velocity perturbations and thus have little impact on the DM perturbations and background expansion rate. For these reasons, we do not consider modifications to the BAO data.

The redshift dependence of the effect of DM-DE scattering can be seen in Fig.~\ref{fig:sig8}. We observe that the impact of $\Gamma_{\rm DMDE}$ is negligible for redshifts above $z\sim 1.5$ and that the equation of state parameter of DE suppresses the impact of DM-DE scattering as it approaches $w=-1$.  This causes a degeneracy when varying both parameters and we thus marginalize over $w$ when deriving constraints on $\Gamma_{\rm DMDE}$.

An important distinction from previous studies on DM-DE scattering from cosmological observables is that we do not make use of late-redshift small-scale weak lensing measurements from galaxies. Indeed, Ref.~\cite{Poulin2023Sigma-8} used a prior on $S_8$ constructed from data from the Dark Energy Survey while Refs.~\cite{Kumar2017ObservationalConstraints,Asghari2019OnStructure} used data from the Canada-France-Hawaii Telescope Lensing Survey~\cite{Heymans2013CFHTLensTomographic}. We instead use CMB gravitational lensing data to $L\sim 1250$ modeled with an adapted non-linear prescription.

It is worth noting that neither of the previous studies cited used a non-linear model in obtaining their constraints. The lensing data from the Dark Energy Survey and the Canada-France-Hawaii Telescope include information on very small scales and at late times where the DM and baryon dynamics are highly non-linear. When using observables on those scales, it is therefore preferable to calibrate models to sufficiently detailed numerical simulations accounting for this added complexity. While there exist simulations of interacting dark sectors~\cite{Baldi2015SimulatingMomentum, Baldi2017StructureFormation}, the redshift evolution of the matter power spectrum from the simulations does not match what we find with the $w\Gamma$CDM model using linear theory. This suggests that the physics implemented in these simulations represents a different type of interaction than what we are modelling.

\subsection{\label{sec:sph-collapse} Non-Linear Model}

In the absence of high-resolution simulations, one can model differences in the non-linear behavior of matter in a model beyond $\Lambda$CDM through the ``spherical-collapse" model~\cite{Asgari2023TheHalo}. This model treats halos as single, spherically symmetric overdensities and is used to calculate two important quantities which enter in the halo model: (i)~the critical overdensity, $\delta_c$, which a region of space must reach to collapse into a bound structure, and (ii)~the virial overdensity, $\Delta_\mathrm{v}$, of the structure after it has undergone collapse. The former is used to calculate the distribution of halos as a function of mass through the halo mass function while the latter gives the virial radius used to calculate halo density profiles. Standard expressions for these quantities assume $\Lambda$CDM and must be modified in the case of non-$\Lambda$CDM evolution. 

To account for the non-linear evolution of the matter density on semi-linear scales, we follow Ref.~\cite{Mead2016HMCODE-Modif} and calculate the critical overdensity and virial overdensity under the spherical collapse approximation while allowing for DM-DE coupling. 
As in previous analyses~\cite{Simpson2010ScatteringOf, Poulin2023Sigma-8}, we neglect the impact of clustering of dark energy perturbations given its high sound speed and focus on DM, which we take to be cold, but we allow for variations in the equation of state of dark energy away from its $\Lambda$CDM value of $w=-1$. In the spherical collapse approximation, we consider a spherical region of constant density  
with radius $r$. 

Assuming that the coupling only affects the rate of change of the DM velocity, we can write the non-linear equation of motion for the DM overdensity as
\begin{align}
    &\ddot{\delta}_{\rm DM} + 2H\left(1+\frac{\Gamma_{\rm DMDE}}{2H\bar{\rho}_{\rm DM}}\right)\dot{\delta}_{\rm DM} -\frac{4}{3}\frac{\dot{\delta}_{\rm DM}^2}{1+\delta_{\rm DM}}\nonumber \\&\;\;\;\;\;= \frac{3}{2}(1+\delta_{\rm DM})H^2\Omega_\mathrm{m}(a) \delta_{\rm DM}, \label{eq:main-nonlinear-delta}
\end{align}
where $\Omega_\mathrm{m}(a)$ is the redshift-dependent mean matter density (we use $\Omega_\mathrm{m}$ to denote the value at the present day). To monitor the evolution of the spherical overdensity, we define the differential radius as
\begin{align}
    q\equiv \frac{r}{r_i} - \frac{a}{a_i},
\end{align}
where $r_i$ and $a_i$ are the initial radius and scale factor at some high redshift (during the matter-dominated era). 
Under the assumption that the mass contained in the overdensity is constant over time, we have that the change in the density contrast depends only on the change in volume giving
\begin{align}
    1+\delta = \left(\frac{1}{qa_i/a+1}\right)^3(1+\delta_i),\label{eq:delta_ic}
\end{align}
where $\delta_i$ is the initial overdensity. Following the procedure used for modified gravity models in Ref.~\cite{Schmidt2009NonlinearEvolution,Herrera2017CalculationOf}, we make the change of variables $y\equiv \ln a$ with Eq.~(\ref{eq:main-nonlinear-delta}) and derive the equation of motion for the differential radius
\begin{widetext}
\begin{align}
    \frac{d^2q}{dy^2}+&\frac{1}{H}\frac{dH}{dy} \frac{dq}{dy} + \frac{\Gamma_{\rm DMDE}}{H\bar{\rho}_{\rm DM}}\left(\frac{dq}{dy}-q\right)  =-\frac{1}{2} \frac{\Omega_{\rm m } a^{-3}+(1+3w) \Omega_{\Lambda }a^{-3(1+w)}}{\Omega_{\rm m } a^{-3}+\Omega_{\Lambda }a^{-3(1+w)}} q -\frac{1}{2} \frac{\Omega_{\rm m } a^{-3}}{\Omega_{\rm m } a^{-3}+\Omega_{\Lambda }a^{-3(1+w)}}\left(\frac{a}{a_i}+q\right) \delta. \label{eq:q-eom}
\end{align}
\end{widetext}
This equation is found by first observing that the change in the radius of the overdensity is related to the matter it contains, so that $\ddot{r}/r \propto \rho_{\rm m}$ in $\Lambda$CDM. One can then make the change of variables $r\to q$ and $a\to y$ and use Eq.~(\ref{eq:delta_ic}) to obtain the density of matter inside the perturbation and derive the full equation for the differential radius $q$ (Eq.~\ref{eq:q-eom}). We recover the equation of motion for $q$ found in the literature in the limit $(\Gamma_{\rm DMDE}, \; w)\to (0, \;-1)$. We solve Eq.~(\ref{eq:q-eom}) numerically with the initial conditions $q_i =0$ and $[dq/dy]_i = -\delta_i/3(1+\delta_i)$. The spherical overdensity collapses when $r=0\Rightarrow q=-a/a_i$ and the maximal radius is attained when $dq/dy=-a/a_i$. After identifying the redshift of collapse $z_c$ we vary the value of the initial overdensity $\delta_i$ until $z_c(\delta_i) =0$ (or any specific redshift at which the matter power spectrum is evaluated). We then take the same $\delta_i$ and calculate the corresponding value of the \emph{linear} overdensity at the collapse redshift using
\begin{align}
    \delta_c = \frac{D(z_c)}{D(z_i)}\delta_i,
\end{align}
where $D$ is the linear growth factor found by solving for the time-dependent part of Eq.~(\ref{eq:main-linear-delta}) on large scales (where the solution is assumed to be separable). This value is known as the critical overdensity and has a reference value of $\delta_c = 1.686$ for a matter-dominated universe. The critical overdensity is the quantity that indicates which regions collapse to form bound structures and is therefore central in the calculation of the halo mass function (and by extension the non-linear matter power spectrum).

The virial overdensity $\Delta_\mathrm{v}$ can also be found using a similar approach. Virialization in the presence of evolving dark energy has a density-profile-dependent effect on halos, which motivates the use of numerical simulations. We leave the treatment of the fully non-linear regime and halo interiors to future studies. We follow the conventions of Ref.~\cite{2021MeadHMCODE-2020} and take the virial radius to be half of the maximal radius.

\subsection{Numerical Implementation}

As stated previously, we adapt the Boltzmann code \texttt{CLASS} to include the effects of the $\Gamma_{\rm DMDE}$ parameter as described by the system of equations Eq.~(\ref{eq:DM-dens-deriv})-(\ref{eq:DE-vel-deriv}). These changes allow us to track the impact of DM-DE interaction in the linear regime. The code \texttt{CLASS} also includes a numerical implementation of the halo model \texttt{HMCode}~\cite{Mead2015HMCodeComputation} in order to incorporate non-linear corrections to the matter power spectrum. This approach is based on the assumption that all the matter in Universe is contained within halos of different masses and that, at a fixed redshift, halos of the same mass have the same density profile. Under this assumption, the distribution of matter on non-linear scales can be found by integrating over the radial density profiles and over halo masses. The latter integral is weighted by the abundance of halos given their mass, a quantity known as the halo mass function. There are further corrections which can be applied to account for changes in density profiles such as baryonic feedback parameters. In all cases, the corrections from the halo model leave the linear power spectrum unchanged at large scales and early times, where linear theory is a good approximation.

We modify the \texttt{HMCode} implementation in \texttt{CLASS} to include the changes to the spherical and virial overdensities  described in Sec.~\ref{sec:sph-collapse} as these two numbers enter in the computation of the shape of the halo mass function and the size of DM halos. Since Eq.~(\ref{eq:q-eom}) would be too computationally expensive to solve for every set of parameters of the non-linear model, we instead solve the system for a finite set of values of $\{ \Omega_{\rm m}, \;w,\;\Gamma_{\rm DMDE} \}$ and calibrate fitting functions for $\delta_c$ and $\Delta_{\rm v}$. For more detail on the fitting function, see Appendix~\ref{app:fitting-func}.

The version of \texttt{HMCode} implemented in \texttt{CLASS} uses a $\delta_c$ value fitted to $N$-body simulations rather than its spherical collapse value~\cite{Mead2016HMCODE-Modif}. In order to modify the spherical overdensity in a self-consistent way, we first update the fitting functions for $\delta_c$ and $\Delta_\mathrm{v}$ in \texttt{CLASS} to match that of the 2020 \texttt{HMCode}~\cite{2021MeadHMCODE-2020}. This gives us an implementation of \texttt{HMCode} which matches more closely the 2020 version of the halo model compared to the 2016 version currently implemented in \texttt{CLASS}\footnote{The 2020 version is not yet implemented in \texttt{CLASS}.}. However, our changes do not include the modifications to baryonic feedback parameters made between the 2016 and the 2020 halo model versions. We then compare the results with another Boltzmann code (\texttt{CAMB}~\cite{Lewis1999CAMB}), which implements the 2020 \texttt{HMCode} version, and verify that the non-linear power spectra for $\Lambda$CDM from both our implementation and the version in \texttt{CAMB} agree at the 2\% level for $k\leq 1 \;h/$Mpc (see Appendix~\ref{app:fitting-func}). On fully non-linear scales, we do not implement the detailed baryonic feedback models of the latest \texttt{HMCode} and this results in relative deviations of a few parts per thousand. This matches the expected variation between non-linear models themselves. The difference between \texttt{HMCode} 2020 from \texttt{CAMB} and our modified non-linear model results in a relative difference in the lensing convergence power spectrum of less than $10^{-3}$ for $L\leq 300$ and $10^{-2}$ for $L\leq 2000$.

By changing the linear matter power spectrum and the virial and spherical overdensities, our model accounts for changes in the two-halo term (the term which accounts for two-point correlations in the matter field between halos) and halo mass function in $w\Gamma$CDM. As shown in detail in Ref.~\cite{2021MeadHMCODE-2020}, these components of the halo model are the most important at scales larger than $k \sim 1\; h/$Mpc. We neglect changes to the halo density profiles and the halo concentration-mass relation, as these would require non-linear simulations beyond the scope of this work. The impact of halo concentration on the non-linear matter power spectrum manifests at scales smaller than $k\sim 2\;h/$Mpc~\cite{2021MeadHMCODE-2020}. From the calculations of Ref.~\cite{Qu2023ACTLensing} based on the shape of the lensing kernel, most of the CMB lensing signal originates from structure around redshift $z\sim 1$-2. Furthermore, the power spectrum analysis of Ref.~\cite{Chabanier2019MatterPower} suggests that scales below $k\sim 0.5\;h$/Mpc contribute most to the CMB lensing signal for our range of $L$. The adaptations made to the halo model should thus be sufficient to capture the change in our choice of observables.

\begin{figure}
    \centering
    \includegraphics[width=\linewidth]{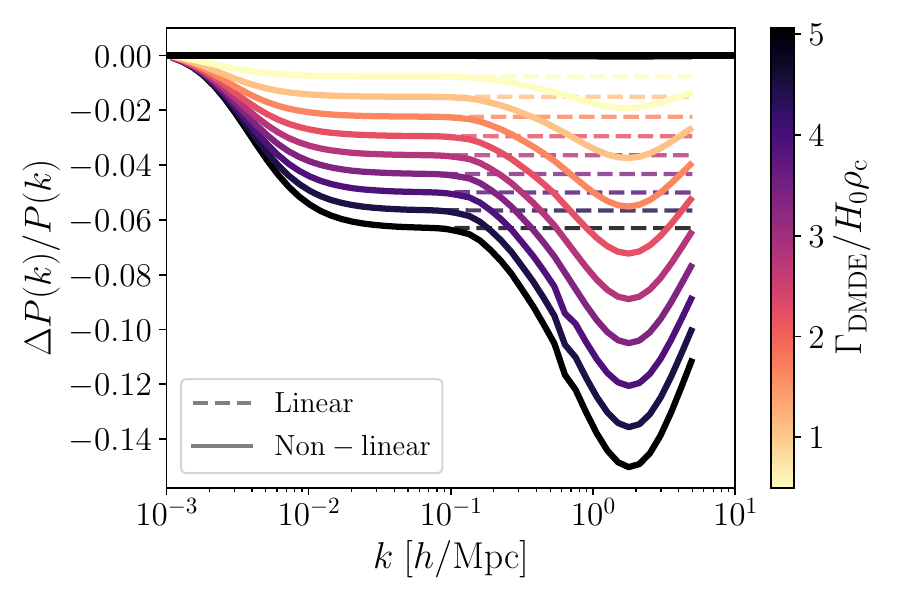}
    \caption{Matter power spectra computed from our halo model and from linear theory at redshift $z=1$ with $w=-0.98$ for various values of $\Gamma_{\rm DMDE}$ compared to the $w$CDM result with $\Gamma_{\rm DMDE}/(H_0\rho_{\rm c})=0$ (to isolate the impact of the scattering term).}
    \label{fig:pks_nl_lin}
\end{figure}

Examples of the impact of $\Gamma_{\rm DMDE}$ on the matter power spectrum are shown in Fig.~\ref{fig:pks_nl_lin}. We observe an increase in suppression with respect to $w$CDM when accounting for non-linearities. This suppression reaches its maximum value around $k\sim 1-2 \,\, h/\mathrm{Mpc}$ and decreases on smaller scales. This characteristic spoon-like feature is present when non-linear structure growth is delayed and the scale of non-linearity increases. On sufficiently small scales, the non-linear growth due to mode-coupling to large-scale modes dominates despite the initial suppression, and the power spectrum can eventually exceed the linear theory prediction. This effect has been found to occur in cosmologies with a scale-dependent power spectrum suppression, such as models with massive neutrinos~\cite{Branbyge2008TheEffect,Viel2010TheEffect,Bird2012MassiveNeutrinos,Liu2018}, ultralight axions~\cite{Vogt2023ImprovedMixed}, and some quintessence models~\cite{Alimi2010ImprintsOf}. A more intuitive explanation for this spoon-shaped ratio is that the change in the matter power spectrum introduces a suppression in the number of high mass halos. Conversely, the number density of low mass halos in $w\Gamma$CDM and $\Lambda$CDM are comparable (this occurs also in cosmologies with massive neutrinos compared to $\Lambda$CDM)~\cite{Massara2014TheHalo}. Since the one-halo term in the power spectrum is dominated by low-mass halos on small scales and by high-mass halos on intermediate scales, we observe a decrease in the power spectrum ratio followed by an increase to match the $\Lambda$CDM result.

\subsection{CMB Lensing Power Spectrum}
The main addition in our datasets is the CMB lensing convergence power spectrum from ACT. The binned datapoints with error bars for the extended $L$ range are shown in Fig.~\ref{fig:lensing_with_Gamma}. Lensing of the CMB is a particularly good probe for constraining $\Gamma_{\rm DMDE}$ as it is sensitive to changes in the growth of the matter power spectrum. Under the Limber approximation~\cite{Limber1953TheAnalysis}, the lensing convergence angular power spectrum of the CMB is given by
\begin{align}
    C_L^{\kappa\kappa} = \int_0^{z_{\star}} dz \frac{H(z)}{\chi^2(z)} W^2(z) P_{\rm NL}\left(k=L/\chi(z), z\right), \label{eq:clkk-theo}
\end{align}
where $P_{\rm NL}$ is the non-linear matter power spectrum, $z_\star$ is the redshift of the last scattering distance, $\chi$ is the comoving distance, $\chi_\star = \chi(z_\star)$ and
\begin{align}
    W(z) = \frac{3\Omega_{\rm m}}{2} \frac{H_0}{H^2(z)} (1+z) \chi(z) \frac{\chi_\star-\chi(z)}{\chi_\star}.
\end{align}
In $w\Gamma$CDM, the DM-DE interaction causes a suppression of the matter power spectrum which appears in the lensing convergence as a reduction in $C_L^{\kappa\kappa}$. This effect is shown for $\Gamma_{\rm DMDE}/(H_0 \rho_c) \leq 10$ in Fig.~\ref{fig:lensing_with_Gamma}. Given the integral over redshift in Eq.~(\ref{eq:clkk-theo}), CMB lensing is also sensitive to the time evolution of the matter power spectrum. To constrain the DM-DE interaction, it is beneficial to include as much information on the small-scale (high $k$) matter power spectrum as possible. This motivates the use of the ACT DR6 lensing data, which extend to $L=1250$, rather than $L=400$ for \textit{Planck} lensing.


\section{\label{sec:data} Data and Methodology}

\begin{figure}[b]
    \centering
    \includegraphics[width=\linewidth]{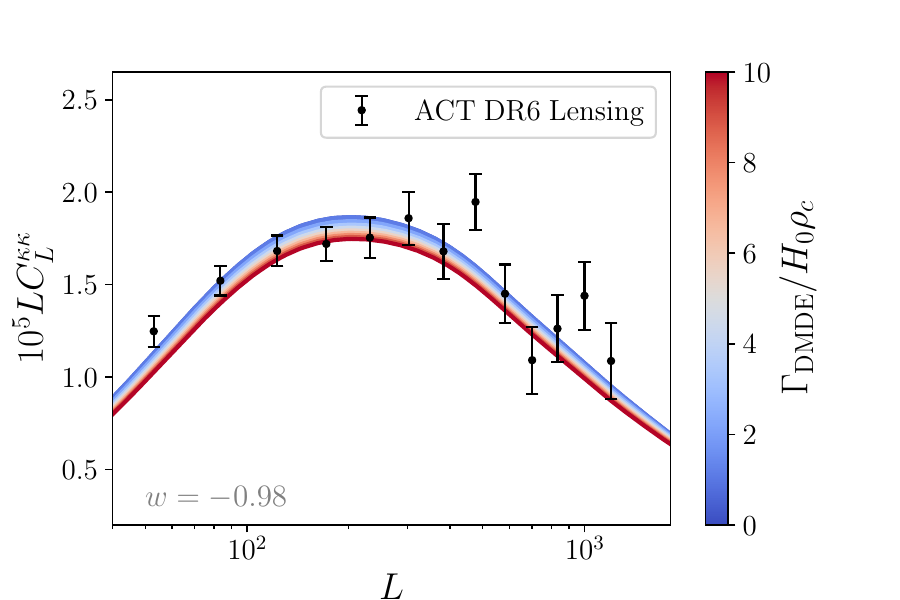}
    \caption{Lensing convergence power spectrum for different values of $\Gamma_{\rm DMDE}$ when the dark energy equation of state is fixed at $w=-0.98$.}
    \label{fig:lensing_with_Gamma}
\end{figure}

To constrain the $w\Gamma$CDM model, we use the following datasets including observations of the cosmic microwave background, baryon acoustic oscillations, and weak gravitational lensing as listed below.
\begin{enumerate}
    \item \textbf{\textit{Planck} CMB 2-point}: From the \textit{Planck} 2020 data release~\cite{Planck2018GravitationalLensing, Planck2018Likelihood}, the HiLLiPoP likelihood~\cite{Couchot2017Hillipop} for high-$\ell$ TTTEEE, and the original and Sroll2 low-$\ell$ TT and EE likelihoods. We find comparable results when using the CamSpec likelihood~\cite{Efstathiou2019ADetailed} for high-$\ell$ (see Appendix~\ref{app:likelihoods}).
    \item \textbf{SDSS+6dFGS BAO}: From the Sloan Digital Sky Survey, we include the DR7 MGS~\cite{DR7MGS}, the DR12 BOSS~\cite{SDSSDR12BOSS}, and the DR16 eBOSS~\cite{SDSSDR16eBOSS} BAO likelihoods including information from the Lyman-$\alpha$ forest. We add to this the results of 6dFGS survey~\cite{Beutler20116dFGalaxy}. 
    \item \textbf{ACT DR6+\textit{Planck} CMB lensing}: We use the updated PR4 lensing likelihood~\cite{Carron2022CMBLensing} on its own and also in combination with the ACT DR6 CMB lensing likelihood~\cite{Qu2023ACTLensing, Madhavacheril2023ACTLensing}. For the ACT lensing data, we use two configurations which are the baseline (\textbf{base.}) with $40 < L < 763$ and the extended (\textbf{ext.}) going to $40 < L< 1250$. The data points and error bars for these data are shown along with the theory curve in Fig.~\ref{fig:lensing_with_Gamma}. We use the version of the ACT DR6 likelihood which also incorporates the lensing data from \textit{Planck}, and therefore all uses of the ACT lensing likelihood include lensing data from \textit{Planck}.
\end{enumerate}

This choice of datasets is very close to the combination used in the ACT DR6 lensing analysis~\cite{Madhavacheril2023ACTLensing} with the exception that the latter used the CamSpec instead of the HiLLiPoP likelihood for the primary CMB and the fact that we have added the quasar BAO data from eBOSS DR16 which extends to redshift $z=2.2$ and the Lyman-$\alpha$ forest which extends to $z=3.5$. One major difference with the analysis of the $w\Gamma$CDM model found in Ref.~\cite{Poulin2023Sigma-8} is that we do not use measurements of the growth rate ($f\sigma_8$) from redshift-space distortions. This is due to the fact that these measurements are based on templates for a $\Lambda$CDM cosmology and may not be suitable for testing models with large deviations from their fiducial model.

We sample the posterior distribution of the cosmological parameters $\left\{\Omega_b h^2, \Omega_c h^2, \tau, \ln 10^{10} A_s, n_s, 100\theta_s\right\}$ which are the standard $\Lambda$CDM parameters (see Ref.~\cite{Planck2018Likelihood}) along with the new model parameters $\left\{w, \Gamma_{\rm DMDE} \right\}$. We impose the same priors on the $\Lambda$CDM parameters as in the analysis of Ref.~\cite{Madhavacheril2023ACTLensing}, but fix the sum of the neutrino masses to $0.06$ eV. The matter power spectrum model calibration in the high neutrino mass regime is particularly dependent on the implementation of neutrinos used in the simulations~\cite{2021MeadHMCODE-2020}. Thus, we leave the modelling of massive neutrinos in the presence of DM-DE scattering in the non-linear regime for future work. The DM-DE scattering also suppresses the matter power spectrum as do massive neutrinos, but the former has a much different scale and redshift dependence as is shown in Fig.~\ref{fig:sig8}. It has been shown in simulations that the effect of massive neutrinos has a different $L$-dependent suppression in the lensing convergence power spectrum, matching the $\Lambda$CDM result more closely on large scales~\cite{Vielzeuf2023DEMNUniThe}. We therefore expect massive neutrinos and $\Gamma_{\rm DMDE}$ to be only marginally degenerate.

For the $w\Gamma$CDM parameters, we use the linear priors $10^{-8} < \Gamma_{\mathrm{DMDE}}/(H_0\rho_{\rm c}) < 50$ and  $-1+10^{-6} < w < -0.6$. These bounds are chosen to ensure the priors are uninformative and to avoid numerical instabilities arising when $1+w = 0$ in the expression for the DE velocity divergence in Eq.~(\ref{eq:DE-vel-deriv}) or $\Gamma_{\rm DMDE} = 0$ in the non-linear fitting functions (see Appendix~\ref{app:fitting-func}). We use the Markov Chain Monte Carlo (MCMC) sampling software \texttt{Cobaya}~\cite{Torrado2021Cobaya} for sampling and the \texttt{GetDist} package~\cite{Lewis2019GetDist} for the analysis and plotting. We run the chains until the Gelman-Rubin statistic reaches $|R-1| \leq 0.02$.


\section{\label{sec:results} Results}
\begin{figure*}[t]
    \centering
    \includegraphics[width=0.65\textwidth]{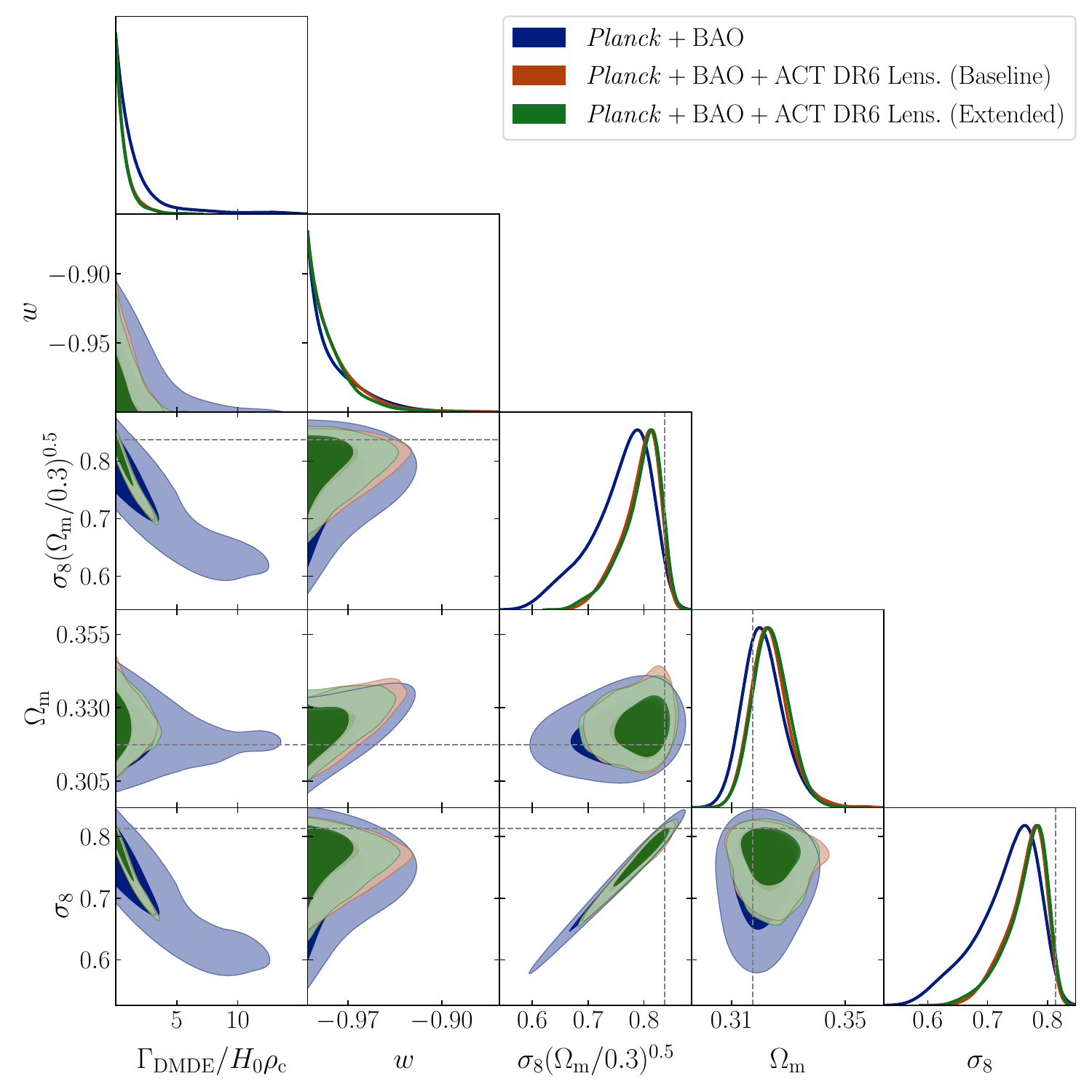}
    \caption{Triangle plot of posterior distributions of $w\Gamma$CDM model parameters along with $S_8$, $\sigma_8$ $\Omega_{\rm m}$ for various datasets. The dashed grey lines denote the best-fit value for the \textit{Planck}+BAO+ACT DR6 Lensing extended data.}
    \label{fig:P18_BAO_ACT}
\end{figure*}
We evaluate the posterior distributions for the $w\Gamma$CDM model parameters for 3 different dataset combinations: 
\begin{enumerate}
\item \textit{Planck} primary CMB and lensing with BAO data from SDSS and 6dFGS,
\item same as (1) with ACT DR6 lensing baseline,
\item same as (1) with ACT DR6 lensing extended.
\end{enumerate}

We display the resulting 68\% and 95\% confidence contours in Fig.~\ref{fig:P18_BAO_ACT} and Table~\ref{tab:results}. We observe a sharp tightening of the constraints when including the new ACT DR6 lensing data. This is due to the fact that this dataset allows us to constrain a larger range of scales of the matter power spectrum~\cite{Qu2023ACTLensing}. In the case where $w\approx -1$, the effects of $\Gamma_\mathrm{DMDE}$ are dampened and the parameter becomes unconstrained. This can be seen in Fig.~\ref{fig:P18_BAO_ACT} as the posterior distribution for $\Gamma_{\rm DMDE}/(H_0\rho_{\rm c})$ widens as $w$ approaches $-1$. This situation is present with many multi-parameter extensions of $\Lambda$CDM where one of the added parameters going to zero (or to the $\Lambda$CDM value) erases completely the effect of varying another added parameter, such as ultralight axions~\cite{Rogers2023UltralightAxions} and early dark energy~\cite{Poulin2023UpsAnd,2023arXiv231019899M}. In Fig.~\ref{fig:P18_BAO_ACT}, we observe a shift between the peak in the posterior distributions of $S_8,\;\sigma_8,\;\Omega_\mathrm{m}$ and their best-fit values. Looking at Table~\ref{tab:comp-wcdm}, maximum-likelihood values for $\Gamma_\mathrm{DMDE}$ and $w$ lie at the edge of their priors. The marginalized distribution of the parameters which are degenerate with these are non-Gaussian, with most of the weight close to the best-fit and a tail extending away from it. The shift between the maximum-likelihood values and the peak of the posterior distribution is indicative of potential prior volume effects. However, the changes in the minimum $\chi^2$ as shown in Table~\ref{tab:comp-wcdm} are very small and the posterior distributions of the parameters are all consistent with $\Lambda$CDM at the 68\% C.L., indicating that such prior volume effects are negligible. It is also worth noting that the differences in $\chi^2$ and best-fit parameters are mostly due to small numerical effects and not physical.

\begin{table*}
\caption{\label{tab:results}68\% C.L. (95\% C.L.) parameter constraints for $w\Gamma$CDM}
\begin{ruledtabular}
\begin{tabular}{lcccccc}
Datasets & $\Gamma_{\rm DMDE}/(H_0\rho_c)$ & $w$ & $S_8$ & $\sigma_8$ & $H_0$ & $\Omega_{\rm m}$ \\
\hline\\
\textit{Planck} + BAO & $< 1.64 \; (< 7.72)$ & $< -0.975$ & $0.758^{+0.067}_{-0.034}$ & $0.733^{+0.064}_{-0.033}$ &$66.65^{+0.71}_{-0.49}$ & $0.3211^{+0.0061}_{-0.0077}$ \\ 
\textit{Planck} + BAO + ACT Lens. Base. & $< 0.861 \; (< 2.48)$ & $< -0.978$ & $0.793^{+0.042}_{-0.022}$ & $0.764^{+0.040}_{-0.021}$ & $66.50^{+0.67}_{-0.42}$ & $0.3235^{+0.0055}_{-0.0073}$ \\
\textit{Planck} + BAO + ACT Lens. Ext. & $< 0.831 \; (< 2.81)$ & $< -0.980$ & $0.795^{+0.043}_{-0.021}$ & $0.765^{+0.042}_{-0.020}$ & $66.50^{+0.59}_{-0.47}$ & $0.3238^{+0.0060}_{-0.0068}$  \\

\end{tabular}
\end{ruledtabular}
\end{table*}

\begin{figure}
    \centering
    \includegraphics[width=0.8\linewidth]{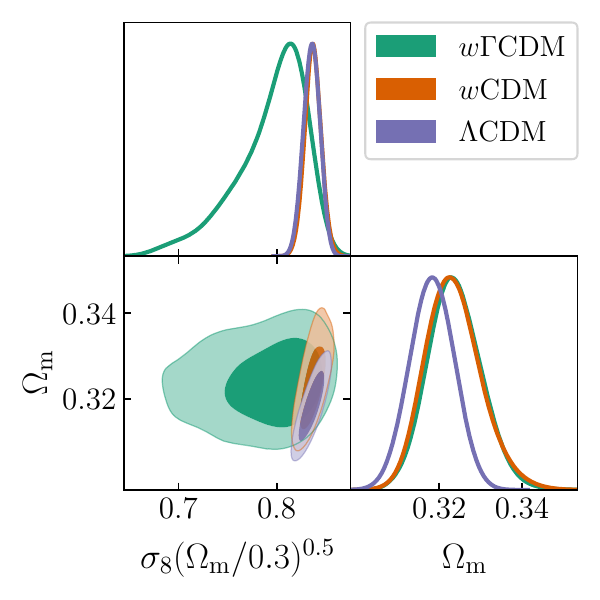}
    \caption{Changes in the $S_8-\Omega_{\rm m}$ posterior distributions when varying $w$ and $\Gamma_{\rm DMDE}$ compared to $\Lambda$CDM for the full dataset including \textit{Planck} primary CMB with PR4 lensing, BAO data from SDSS and 6dFGS, and ACT DR6 lensing (extended).
    }
    \label{fig:S8-model-comparison}
\end{figure}

In Fig.~\ref{fig:S8-model-comparison}, we compare the preferred values of $S_8$ and $\Omega_{\rm m}$ obtained from the $\Lambda$CDM, $w$CDM, and $w\Gamma$CDM models using the full dataset including \textit{Planck} primary CMB with PR4 lensing, BAO data from SDSS and 6dFGS, and ACT DR6 lensing (extended). We note that varying $w$ away from $-1$ does not change the inferred value of $S_8$ on its own, but widens the posterior on $\Omega_{\rm m}$. This comes from the fact that both the DE equation of state and the amount of matter in the Universe affect the cosmic expansion rate during the matter-dominated era, which introduces a degeneracy between the two parameters. However, we only observe a widening in the $S_8$-direction when varying $\Gamma_{\rm DMDE}$ since this parameter changes the shape of the matter power spectrum. Marginalizing over this parameter gives a value of $S_8$ compatible with the results from galaxy weak lensing surveys (see Sec.~\ref{sec:intro}).

\begin{figure}
    \centering
    \includegraphics[width=0.8\linewidth]{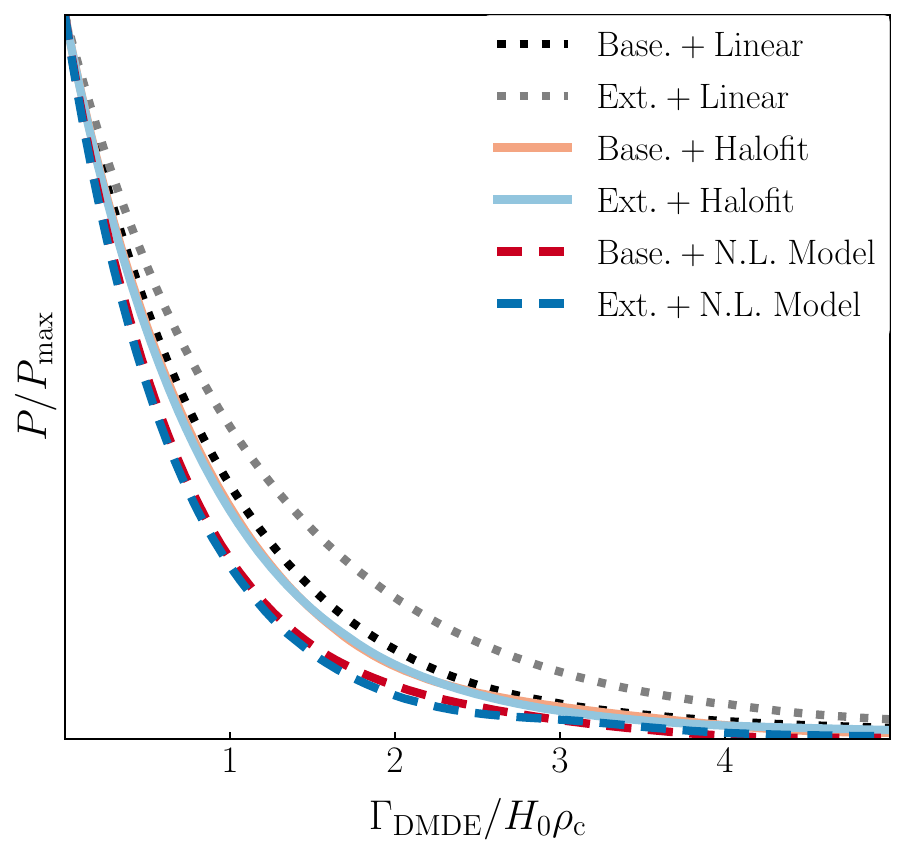}
    \caption{Comparison of posterior distribution in $\Gamma_{\rm DMDE}$ for different choices of non-linear model using \textit{Planck} + BAO + ACT DR6 baseline or extended lensing.}
    \label{fig:non-lin-comparison}
\end{figure}

\begin{table*}
\caption{\label{tab:comp-wcdm} Best-fit values comparison using CMB + BAO + CMB Lensing (Ext.). The best-fit values of some parameters lie at the edge of their priors and the slight changes in $\chi^2$ are due to numerical effects.
}
\begin{ruledtabular}
\begin{tabular}{lcccccc}
Model & $\Gamma_{\rm DMDE}/(H_0\rho_c)$ & $w$ & $S_8$ & $\sigma_8$ & $\Omega_{\rm m}$ & $\Delta \chi^2(\Lambda\mathrm{CDM})$\\
\hline\\
$w\Gamma$CDM & $10^{-8}$ & $-1+10^{-6}$ & $0.837$ & $0.813$ &$0.3175$ & $-0.48$\\
$w$CDM & -- & $-1+10^{-6}$ & $0.836$ & $0.812$& $0.3185$ & $-0.39$\\
$\Lambda$CDM & -- & -- & $0.837$ & $0.811$ & $0.3193$ & $0.0$ \\
\end{tabular}
\end{ruledtabular}
\end{table*}

To assess the impact of our non-linear model on the constraints on $\Gamma_{\rm DMDE}$, we compare the constraints on $\Gamma_{\rm DMDE}$ using the dataset combination CMB 2-point + BAO + CMB lensing (extended) with three different non-linear treatments. The first is using linear theory, the second is using the \texttt{Halofit} prescription~\cite{Peacock2014HALOFIT}, and the last is our full non-linear model. \texttt{Halofit} is a simulation-calibrated fitting function designed to infer the non-linear matter power spectrum from the shape of its linear counterpart. It has been heavily tested for $\Lambda$CDM cosmologies, but does not account for the impact of the drag force on halo formation. Furthermore, \texttt{Halofit} was found to output a slightly enhanced non-linear spectrum compared to the last iteration of \texttt{HMCode}~\cite{2021MeadHMCODE-2020}. Since the main feature of the DM-DE model is a suppression of the power spectrum, we expect \texttt{Halofit} to under-estimate the impact of $\Gamma_\mathrm{DMDE}$ in the non-linear regime. For this reason, we expect to find a less constraining bound on $\Gamma_{\rm DMDE}$ using this model. It is worth noting however that \texttt{Halofit} is known to break down for cosmological models with a sharp matter power spectrum cut-off such as cosmologies with ultralight particles~\cite{Hlozek2017FutureCMB}. The posterior contours of this comparison are shown in Fig.~\ref{fig:non-lin-comparison} for the baseline and extended ACT DR6 lensing datasets. We find the 68\% confidence level (C.L.) constraints $\Gamma_\mathrm{DMDE} / (H_0 \rho_\mathrm{c}) < 0.831$, $1.03$, $1.27$ using the non-linear, \texttt{Halofit} and linear models respectively, with the full dataset including the extended $L$-range for lensing. Thus, the inclusion of the non-linear corrections improves our constraints by about 35\%. Comparing this with the first row of Table~\ref{tab:results}, this indicates that most of the constraining power in our analysis comes from the CMB lensing data from ACT. While the improvement is small over \texttt{Halofit}, the non-linear model accounting for changes in the spherical overdensity yields the tightest constraints, as expected. Our constraints are also more robust as they are based on a physically-motivated model which includes linear dynamics corrections on semi-linear scales. Furthermore, we exclude data highly sensitive to the clustering of matter on very small scales which would require numerical simulations to model accurately.


\section{Discussion\label{sec:discussion}}
We have performed a new analysis of the $w\Gamma$CDM model with the most recent gravitational lensing data from the Atacama Cosmology Telescope. Unlike previous studies of this model, our approach did not make use of small-scale galaxy weak lensing data and involved a physically motivated halo model for quasi-linear scales. We note that using a prior on $S_8$ as done in Ref.~\cite{Poulin2023Sigma-8} from galaxy weak lensing involves collapsing rich non-linear physics into a prior fitted using linear theory (even if we use our halo model, the prior is on $\sigma_8$ which is based on the shape of the \emph{linear} matter power spectrum). Furthermore, the linear model used to derive this prior is based on $\Lambda$CDM while we consider an extension of this model with significant deviations in the shape of the matter power spectrum. Since the shape of the power spectrum deviates from the linear theory prediction on the scales relevant for galaxy weak lensing, we do not make use of such a prior in our analysis. Incorporating galaxy weak lensing data would require combining such datasets at the likelihood level after either applying scale-cuts that restrict to quasi-linear scales, or after validating the halo model approach with numerical simulations.

We have adapted the Boltzmann code \texttt{CLASS} and its accompanying \texttt{HMCode} implementation. For the latter, we have modified the equations for calculating the spherical collapse and virial overdensities following the methods used for modified gravity models. These modifications have allowed us to arrive at the first non-linear corrections for the $w\Gamma$CDM model. We have integrated these equations numerically and derived fitting functions for their solutions for fast computation. Our model reaches the tightest constraints on strength of the kinetic coupling between DM and DE to date at $\Gamma_{\rm DMDE}/(H_0\rho_{\rm c}) <0.831$. The Bayesian model comparison we have conducted finds no preference for non-zero $\Gamma$, with only a slightly decreased $\chi^2$ when compared to $\Lambda$CDM. The bound we find on this parameter is compatible with the central value found in Ref.~\cite{Poulin2023Sigma-8} which included a prior on $S_8$ from galaxy weak lensing surveys, but we do not find evidence for $\Gamma_{\rm DMDE}>0$.

\begin{acknowledgments}
The authors would like to thank Bruce Partridge, Crist\'obal Sif\'on, Vivian Poulin and Oliver Philcox for useful discussions. AL acknowledges support from NASA grant 21-ATP21-0145. FMcC acknowledges support from the European Research Council (ERC) under the European Union’s Horizon 2020 research and innovation programme (Grant agreement No. 851274). MM acknowledges support from NSF grants AST-2307727 and  AST-2153201 and NASA grant 21-ATP21-0145. JCH acknowledges support from NSF grant AST-2108536, NASA grants 21-ATP21-0129 and 22-ADAP22-0145, the Sloan Foundation, and the Simons Foundation. Computations were performed on the Niagara supercomputer at the SciNet HPC Consortium. SciNet is funded by Innovation, Science and Economic Development Canada; the Digital Research Alliance of Canada; the Ontario Research Fund: Research Excellence; and the University of Toronto~\cite{Scinet,Niagara}. The Flatiron Institute is supported by the Simons Foundation.
\end{acknowledgments}

\appendix

\section{\label{app:fitting-func} Fitting Functions for $\delta_c$ and $\Delta_\mathrm{v}$}

We construct fitting functions for the critical and virial overdensities as a function of $\Omega_\mathrm{m},\;w,\;\Gamma_{\rm DMDE}$. When compared to the numerical solution found by integrating Eq.~(\ref{eq:q-eom}) these give estimates to better than 2\% and 9\% accuracy for  $\delta_c$, $\Delta_{\rm v}$ and read

\begin{widetext}
\begin{align}
    \delta_c &= \frac{1}{X_c}\left[ 1.686 +\sum_{k=1}^3 a_{\mathrm{m}, k} \left(1- \Omega_\mathrm{m}(z)\right)^k +\sum_{n=1}^3 a_{\Gamma, n} \left(\frac{\Gamma_\mathrm{DMDE}}{H_0\rho_c} \right)^n\right],
\end{align}
\begin{align}
    X_c \equiv p_0  \left(\frac{\Gamma_\mathrm{DMDE}}{H_0\rho_c} \right)^{p_1} \left\{1+p_2 \left[\log_{10}\left(\frac{\Gamma_\mathrm{DMDE}}{H_0\rho_c\Omega_\mathrm{m}(z)}\right)\right]^2 +p_3 \log_{10}\left(\frac{\Gamma_\mathrm{DMDE}}{H_0\rho_c\Omega_\mathrm{m}(z)}\right)  \right\} \Omega_\mathrm{m}(z)^{p_4} \left[1+\frac{1}{(1+z)^4}\right],
\end{align}

\begin{align}
    \Delta_\mathrm{v} &= \frac{1}{X_\mathrm{v}}\left[ 178 \Omega_\mathrm{m}(z)^{a_\mathrm{v}}+ b_\mathrm{v} \left(\frac{\Gamma_\mathrm{DMDE}}{H_0\rho_c} \right)^{c_\mathrm{v}} \right],
\end{align}
\begin{align}
    X_\mathrm{v} \equiv 1+q_0  \left(\frac{\Gamma_\mathrm{DMDE}}{H_0\rho_c} \right)^{q_1} \left\{1+q_2 \left[\log_{10}\left(\frac{\Omega_\mathrm{m}(z)}{0.3}\right)\right]^2 +q_3 \log_{10}\left(\frac{\Omega_\mathrm{m}(z)}{0.3}\right)  \right\}\left(\frac{\Omega_\mathrm{m}(z)}{0.3}\right)^{q_4} \left|\log_{10}\frac{\Gamma_\mathrm{DMDE}}{H_0\rho_c}\right|^{q_5} e^{-5(1+z)},
\end{align}

\end{widetext}

where the parameters for $\delta_c$ and $\Delta_\mathrm{v}$ are listed in Table~\ref{tab:fit-params-c} and Table~\ref{tab:fit-params-v}, respectively.
\begin{table}
\begin{ruledtabular}
    \caption{\label{tab:fit-params-c} Fitting function parameters for $\delta_c$}
    \begin{tabular}{c|c}
        Fitted parameter & Numerical value \\ 
        \hline \\
       $a_\mathrm{m, 1}$  & $-0.046$ \\
       $a_\mathrm{m, 2}$  & $0.034$ \\
       $a_\mathrm{m, 3}$  & $-0.018$ \\
       $\log_{10} a_{\Gamma, 1}$  & $-4.26(1+z)^{0.461}$ \\
       $\log_{10} a_{\Gamma, 2}$  & $-2.51(1+z)^{0.683}$ \\
       $\log_{10} a_{\Gamma, 3}$  & $-1.08(1+z)^{0.915}$ \\
       $p_0$ & $0.0284$ \\
       $p_1$ & $1.50$ \\
       $p_2$ & $0.456$ \\
       $p_3$ & $-0.921$ \\
       $p_4$ & $0.638$ \\
         & 
    \end{tabular}
\end{ruledtabular}
\end{table}
\begin{table}
    \begin{ruledtabular}
    \caption{\label{tab:fit-params-v} Fitting function parameters for $\Delta_\mathrm{v}$}
    \begin{tabular}{c|c}
         \hspace{0.75cm}Fitted parameter\hspace{0.75cm} & Numerical value\hspace{0.75cm}  \\ 
        \hline \\
        $a_\mathrm{v}$ & $-0.352$ \\
       $\log_{10} b_\mathrm{v}$ & $1.85(1+z)^{-2.15}$ \\
       $c_\mathrm{v}$ & $0.96$ \\
       $q_0$ & $0.0159$ \\
        $q_1$ & $0.574$ \\
        $q_2$ & $-18.8$ \\
        $q_3$ & $42.9$ \\
        $q_4$ & $0.548$ \\
        $q_5$ & $0.196$ \\
    \end{tabular}
    \end{ruledtabular}
\end{table}

\begin{figure}
    \centering
    \includegraphics[width=\linewidth]{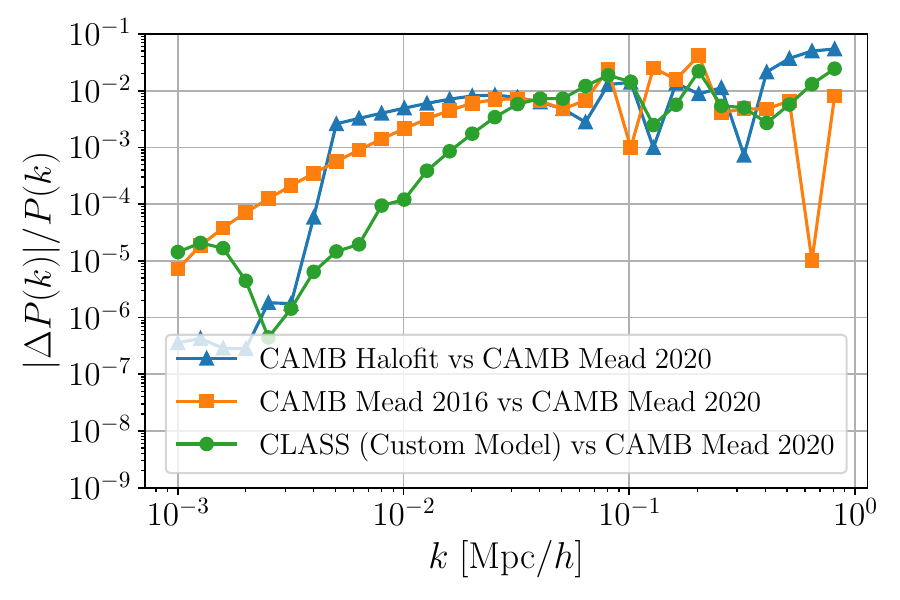}
    \caption{Relative difference in the matter power spectrum (at redshift $z=0$) from different implementations of the non-linear model.}
    \label{fig:pk_errors}
\end{figure}

\begin{figure}
    \centering
    \includegraphics[width=\linewidth]{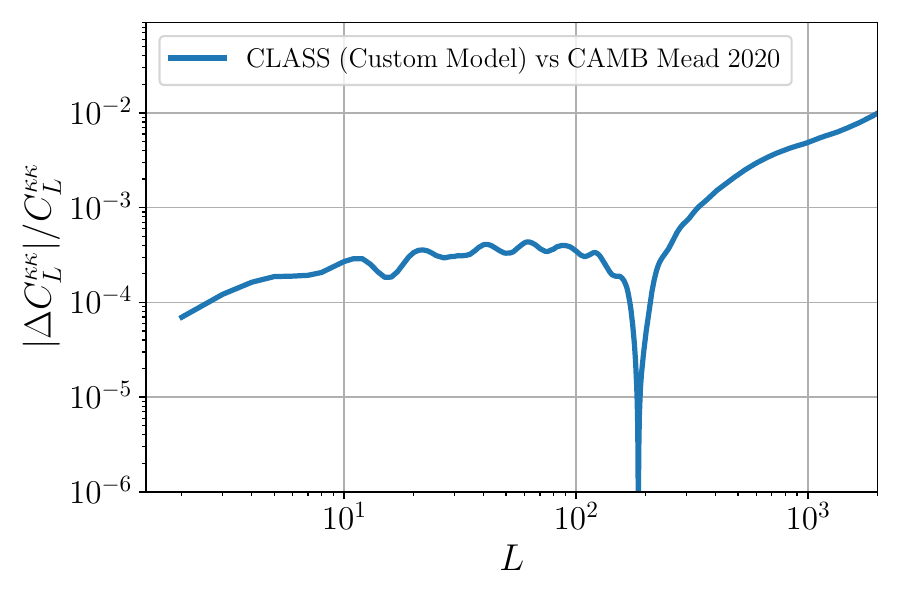}
    \caption{Relative difference in the lensing convergence power spectrum from different implementations of the \texttt{HMCode} 2020 prescription.}
    \label{fig:clkk_errors}
\end{figure}

Our implementation of the non-linear model uses the physical definition of the spherical overdensity, but does not include modelling of baryonic feedback and other small scale corrections present in \texttt{HMCode} 2020. For this reason, our adapted halo model falls in-between \texttt{HMCode} 2020 and its predecessor \texttt{HMCode} 2016. Thus, in the $\Lambda$CDM case, we expect to reproduce the results of \texttt{HMCode} 2020 very well on large scales and to percent-level accuracy for intermediate scales $k\lesssim 0.3h/$Mpc. We display the relative difference between our implementation and the one present in the Boltzmann code \texttt{CAMB} in Fig.~\ref{fig:pk_errors}. In Fig.~\ref{fig:pk_errors} we also compare other halo models within \texttt{CAMB} between each other to verify that the differences induced by our custom model to the reference \texttt{HMCode} are comparable to a simple change in non-linear models. We also integrate the power spectrum using the same Limber approximation and compare the lensing convergence spectra to assess the impact of the above-mentioned variation on our observable. The result is shown in Fig.~\ref{fig:clkk_errors}. We can see that our approach matches the output of \texttt{HMCode} 2020 to better than one part in a thousand for most $L$ with a slight increase in deviation for the largest $L$ considered. This change in the lensing spectrum does not affect our results qualitatively.

\section{\label{app:likelihoods} CMB Likelihoods and $S_8$ Prior}
\begin{figure}
    \centering
    \includegraphics[width=0.7\linewidth]{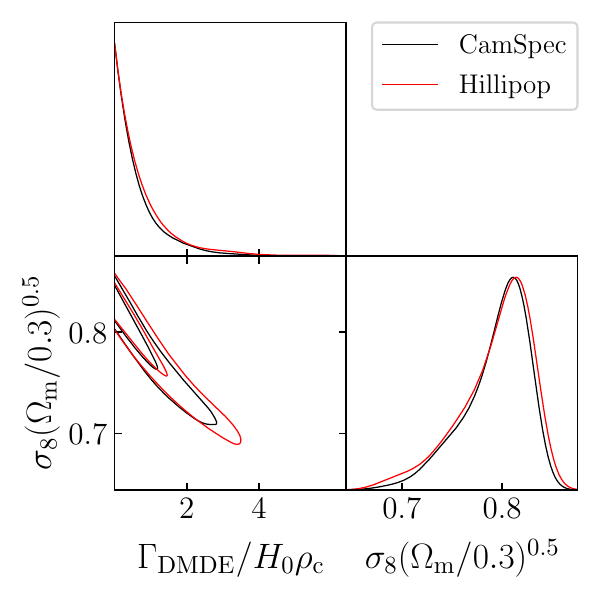}
    \caption{Comparison of the marginalized posterior distributions for $\Gamma_{\rm DMDE}$ and $S_8$ for different choices of the high-$\ell$ primary CMB likelihood.}
    \label{fig:likelihoods}
\end{figure}

In our main analysis we used the HiLLiPoP implementation of the CMB 2-point analysis. We also run our analysis using lensing data from ACT DR6 (extended) and \textit{Planck} PR4 with BAO information from SDSS and 6dFGS while changing the primary CMB likelihood. When switching to the CamSpec likelihood, we find comparable, but slightly stronger constraints with $\Gamma_{\rm DMDE}/(H_0\rho_{\rm c})<0.73\;(2.17)$ at the 68\% (95\%) confidence level. We opt to present the more conservative estimates for the constraints as our main results. The posterior distributions for $\Gamma_{\rm DMDE}$ and $S_8$ derived using either likelihoods are shown in Fig.~\ref{fig:likelihoods}.

\begin{figure}
    \centering
    \includegraphics[width=0.7\linewidth]{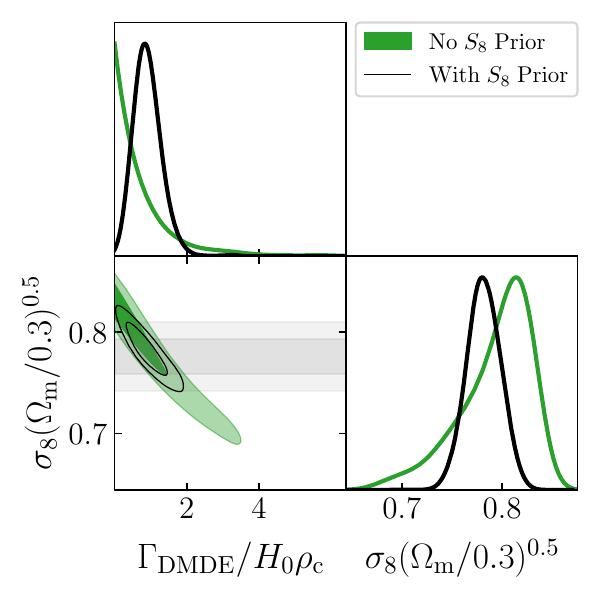}
    \caption{Impact of adding a Gaussian prior on $S_8$ from DES-Y3 for our full dataset \textit{Planck}+BAO+ACT DR6 Extended.}
    \label{fig:S8_prior}
\end{figure}

Finally, we consider the implications of adding a prior on $S_8$ following the results of the DES Y3 constraints from the 3x2pt analysis ($S_8 = 0.776 \pm 0.017$)~\cite{DES2022YR3Params}. After imposing this prior along with the full dataset of \textit{Planck}+BAO+ACT DR6, we find a preference for non-zero $\Gamma_{\rm DMDE}$ at the $\sim 2\sigma$ level with $\Gamma_{\rm DMDE}/(H_0\rho_{\rm c})=0.92^{+0.31}_{-0.44}$. When using this prior with Planck and BAO alone, we find only a preference at the order of $\sim 1\sigma$ with $\Gamma_\mathrm{DMDE} / (H_0 \rho_\mathrm{c}) = 0.943^{+0.065}_{-0.90}$. Such a preference was also observed in Ref.~\cite{Poulin2023Sigma-8}, albeit at the $\sim 3\sigma$ level. We believe this discrepancy is due to the absence of $f\sigma_8$ measurements in our analysis as these data show a preference for slowed growth of structure at low redshift~\cite{Nguyen2023EvidenceFor} and thus favor $\Gamma_{\rm DMDE}>0$. As stated in Sec.~\ref{sec:discussion}, we do not make use of a prior on $S_8$ from galaxy weak lensing in our main results. Nevertheless, for completeness, we display the posterior distribution of $\Gamma_{\rm DMDE}$ with the full dataset along with $S_8$ prior in Fig.~\ref{fig:S8_prior}. \\

\bibliography{apssamp}

\end{document}